\documentclass[journal=jacsat,manuscript=article]{achemso}

\usepackage[version=3]{mhchem}

\usepackage{amssymb}
\usepackage{amsmath}
\usepackage{siunitx}
\usepackage{array,ragged2e}
\newcolumntype{P}[1]{>{\RaggedRight\arraybackslash}p{#1}}
\usepackage{booktabs,siunitx,array,threeparttable,pdflscape}
\usepackage{amsmath}
\usepackage{graphicx}
\usepackage[colorlinks=true, allcolors=blue]{hyperref}
\usepackage[table]{xcolor}
\definecolor{lightgray}{gray}{0.9}
\usepackage{tabularx}
\usepackage{makecell}
\usepackage{multirow, multicol}
\usepackage[shortcuts]{extdash}

\author{James Michels}
\affiliation[UM-CS]
{Department of Computer Science, University of Mississippi, University, MS}

\author{Ramya Bandarupalli}
\affiliation[UM-Pharma]
{Department of BioMolecular Sciences, School of Pharmacy, University of Mississippi, University, MS}

\author{Amin Ahangar Akbari}
\affiliation[UM-Pharma]
{Department of BioMolecular Sciences, School of Pharmacy, University of Mississippi, University, MS}

\author{Thai Le}
\affiliation[Indiana]
{Department of Computer Science, Indiana University, Bloomington, IN}

\author{Hong~Xiao}
\email{hxiao1@olemiss.edu}
\affiliation[UM-CS]
{Department of Computer Science, University of Mississippi, University, MS}

\author{Jing Li}
\email{jli15@olemiss.edu}
\affiliation[UM-Pharma]
{Department of BioMolecular Sciences, School of Pharmacy, University of Mississippi, University, MS}

\author{Erik F. Y. Hom}
\email{erik@fyhom.com}
\affiliation[UM-Bio]
{Department of Biology and Center for Biodiversity and Conservation Research, University of Mississippi, University, MS}

\title{Natural Language Processing Methods for the Study of Protein\=/Ligand Interactions}

\begin{document}

\begin{abstract}
Natural Language Processing (NLP) has revolutionized the way computers are used to study and interact with human languages and is increasingly influential in the study of protein and ligand binding, which is critical for drug discovery and development. This review examines how NLP techniques have been adapted to decode the “language” of proteins and small molecule ligands to predict protein\=/ligand interactions (PLIs). We discuss how methods such as long short-term memory (LSTM) networks, transformers, and attention mechanisms can leverage different protein and ligand data types to identify potential interaction patterns. Significant challenges are highlighted, including the scarcity of high-quality negative data, difficulties in interpreting model decisions, and sampling biases of existing datasets. We argue that focusing on improving data quality, enhancing model robustness, and fostering both collaboration and competition could catalyze future advances in machine-learning-based predictions of PLIs.

\end{abstract}

\section{1. Introduction}
Proteins play a pivotal role as molecular machines essential for biological function. Their functionality often depends on site-specific binding interactions with small molecule ligands that cannot be studied within the protein\=/protein interaction framework~\cite{Sarkar2019-yt}. Understanding such protein\=/ligand interactions (PLIs) is central to drug discovery and development~\cite{Huang2010-yp, Chaires2008-pj} as well as protein engineering efforts~\cite{Woodley2013-cs, Barbosa2018-pq}. Although laboratory experimentation is the traditional approach to studying PLIs and generating "ground\=/truth" data for a specific protein\=/ligand pair, it is both costly and time-consuming, often taking weeks to months~\cite{Vajda2006-vb}. Computational approaches that simulate the underlying physics and chemistry of PLIs such as molecular docking~\cite{Sousa2013-pt} or dynamics simulations~\cite{morris2021-jf} can be less resource intensive but nevertheless demand significant computational and time investment~\cite{Lecina2017-ox}. 

In recent years, machine learning (ML) has provided new avenues for analyzing biological data, leveraging statistical and algorithmic techniques to distill potentially human\=/interpretable insights with little manual intervention. ML models have successfully predicted various protein and molecular attributes~\cite{Ikebata2017-pu, Rives2021-lx, Cao2021-nk, Wang2019-ao, Chithrananda2020-iw}, and have moved us closer to solving the protein folding problem~\cite{Jumper2021-xa, Abramson2024-qh}. As the excitement for ML use in the biological sciences grows, the prediction of protein\=/ligand interactions appears increasingly possible given recent advances in both ML and Natural Language Processing (NLP) \cite{Zhou2020-fe, Patwardhan2023-ui}, the computational study of language~\cite{Bijral2022-ir}. 

\subsection{1.1. Overview of Natural Language Processing (NLP)}
NLP centers on the computational analysis and manipulation of language constructs to bridge the gap between human communication and computer automation. NLP has experienced significant recent breakthroughs as demonstrated by the proliferation of widely used chatbots such as OpenAI’s ChatGPT~\cite{Ray2023-er, Radford_undated-we}, Anthropic’s Claude~\cite{2023-rq}, and Microsoft’s Bing Copilot~\cite{noauthor_undated-ls}. NLP has been further used to summarize texts, deduce author sentiment, solve symbolic math problems, and even generate programming code~\cite{Rahul2020-po, Nasukawa2003-lw, Lample2019-rx, Feng2020-pm}. The effectiveness of NLP is predicated on (human) languages having a structured symbolic syntax and set of rules to assemble basic units known as "tokens" (e.g., characters, words, or punctuation) to form higher\=/order constructs such as sentences or paragraphs. The structured outputs of such a system reflect the grammar, conventions, and styles of the associated language. In NLP, tokens are transformed to encode "meanings" through mathematical vectors such that tokens of similar meaning are positioned closer together in the representational vector space. By analyzing a large collection of data, NLP methods aim to infer emergent relationships between tokens that define the “rules” of a language. Importantly, this inferred set of rules can then be used to perform predictive tasks such as separating tokens into categories, translating text from one language to another, and even predicting whether a literary work will see commercial success~\cite{AshokUnknown-ka, Barbera2021-rh, Wang2022-pi}. 

In the biological domain, NLP methods have been used for a variety of predictive tasks, including inferring disease\=/gene associations~\cite{Bhasuran2018-bp}, predicting tumor gene expression patterns~\cite{Pang2021-xf}, and assigning functional annotations to various protein\=/coding genes~\cite{Cao2021-nk}. More recently, NLP has been applied with unprecedented success in DeepMind's AlphaFold algorithm to predict three\=/dimensional protein structures given only protein sequence data ~\cite{Jumper2021-xa, Abramson2024-qh}. Despite impressive advances, the creation of these NLP models is associated with a sizable computational burden  (see for example,~\cite{Bouatta2021-yz, Skolnick2021-qu, Adadi2018-nl, Box1976-kq, Geirhos2020-lh, Outeiral2022-tf}) and it remains a challenge to understand what and which specific features of the input sequence data fuel predictive success. 

Below, we review contemporary NLP methods as they have been applied in the study of PLIs in recent years. We first describe the relationship between common protein and ligand text representations vis\=/\'a\=/vis the characteristics of human language. We then discuss the dominant NLP\=/based approaches used to study PLIs and provide a comprehensive overview of the diversity of ML models used in this space. We conclude with reflections on remaining challenges in the field and areas that merit future development. 

\section{2. The "Language" of Proteins}
Protein sequences are akin to human language in that they possess a hierarchical order of construction and embody embedded information (Fig. \ref{fig:1}). Human language text is inherently ordered with characters of an alphabet assembled linearly and grouped into words, phrases, and sentences that convey an emergent message. Protein sequences similarly obey a hierarchy of assembly, with amino acids (AAs) serving as the alphabet. When AAs are strung together, secondary structural motifs, domains, and quaternary (multi\=/domain\=/interacting) structures may emerge with properties that contribute to function~\cite{Ferruz2022-pm, Ofer2021-nr}. While external factors such as post-translational modifications and cellular state can play a substantial role in dictating protein three\=/dimensional structure and function, the AA sequence represents the essential blueprint that ontologically defines the properties of a protein~\cite{Ptitsyn1991-jk, Yu2019-mx,Petsko2003-be}. This fact has served as the foundation for bioinformatic analysis of proteins~\cite{Shenoy2010-ye}. Individual AAs and common subsequences contribute to the “information” of the overall protein just as words contribute to the meaning of a text. 

\section{3. The "Language" of Ligands}
The chemical structures of molecules can be similarly translated into text\=/based notations and analyzed computationally~\cite{Garfield1961-xn}. However, unlike the elements of human text and protein sequences, the chemical connectivity patterns of small molecule ligands are not one\=/dimensional. Nevertheless, text\=/based schema has been developed to represent chemical information in a manner convenient for computational analysis~\cite{Wigh2022-om}, with the Simplified Molecular\=/Input Line\=/Entry System (SMILES) format being one of the most widely used~\cite{Weininger1988-da}. 

\begin{figure}[t]
   \centering      
   \includegraphics[scale=0.65]{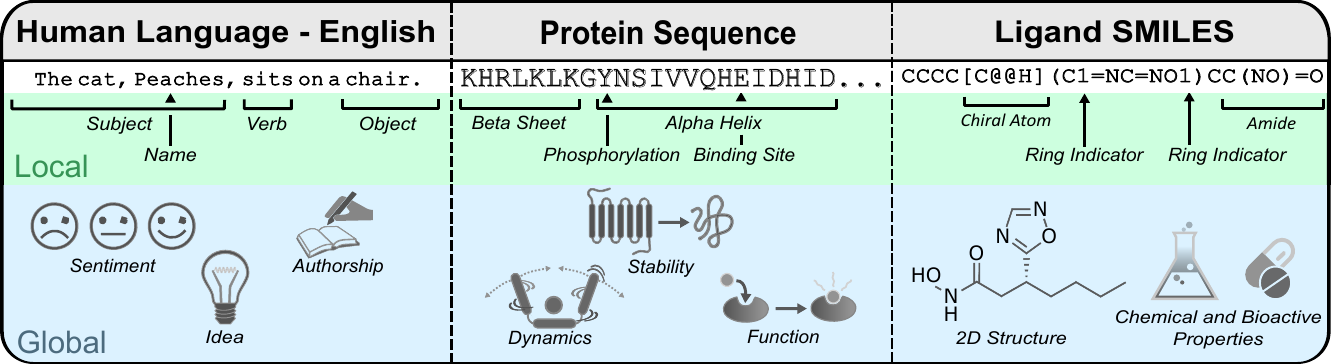}     
 \caption{The Language of Protein Sequences and SMILES: NLP methods can be applied to text representations to infer local and global properties of human language, proteins, and molecules alike. Local properties are inferred characteristics of sub-sequences in text: (i) for a human language, this can include part of speech or a role a specific word serves; (ii) for a protein sequence, this can include secondary structures, post-translational modifications, and functional sites; (iii) for a SMILES string, this can include functional groups and characters used within SMILES syntax to indicate chemical attributes. Global properties are inferred from a text in its entirety: (i) for a human language, this can include information such as authorship, tone, and synopses; (ii) for a protein sequence, this can include the protein's structure, stability, and dynamic properties; and (iii) for a SMILES string, this can include the ligand's 2D molecular structure and other biochemical properties.}
 \label{fig:1}
\end{figure}

SMILES strings are text representations constructed over a depth\=/first traversal of a two\=/dimensional molecular graph (Fig. \ref{fig:1}), with atoms, atomic properties, bonds, and structural properties represented by characters following an established set of conversion rules. Given the memory\=/efficient and somewhat human\=/readable format of SMILES, it has become a standard in chemical databases and computational tools~\cite{Wang2009-en, Degtyarenko2008-od, Wishart2008-ar} and the most commonly used text representation in PLI studies. Although SMILES lacks an intuitive way to determine a chemical equivalent of a “word”, there is a well\=/defined grammar to denote properties and substructures of a molecule. Moreover, the same molecule can be represented by multiple different SMILES strings~\cite{Weininger1988-da}, which is similar to how there could be multiple sentence constructions to convey the same idea in human languages. In NLP applications, incorporating tokens with the same meaning into the training process can yield a robust predictive model~\cite{Wang2021-vu}. The use of multiple SMILES per molecule has been leveraged to guide ML models to discern which parts of a ligand contribute to drug potency~\cite{Bjerrum2017-mr}. 

\section{4. Protein\=/Ligand Interaction Data and Datasets}
Protein\=/ligand binding is a complex process dictated by many factors including protein states, hydrophobicity/hydrophilicity, and conformational flexibility~\cite{Gohlke2012-pd}. The question of \textit{how} to represent a protein and ligand in a computational space is critical and multifaceted. A wealth of information has been collected experimentally and generated through simulation studies on the properties of proteins and ligands, but these data are highly variable with regard to type, quality, and quantity. This section catalogs several primary data representations used in PLI studies. We also discuss the availability, selection, and curation of available data for machine\=/learning\=/based training and evaluation. 

Protein and ligand representations are typically sequence- or structure\=/based. Unlike sequence\=/based text formats, structure\=/based information can appear in multiple forms, e.g., atomic coordinates of protein\=/ligand complexes or contact maps. Some structural information can be artificially reconstructed from sequence\=/based formats through algorithms such as AlphaFold for proteins~\cite{Jumper2021-xa} and RDKit for ligands~\cite{Landrum2013-dh}. PLI studies using machine\=/learning methods will typically select either sequence\=/based or structure\=/based inputs, although there is a growing use of mixed input data types~\cite{Mukherjee2022-xg, Chen2020-ns}. For example, a mixed-data study may represent proteins by AA sequences but ligands by atomic coordinates, a choice based in part on the fact that highly accurate 3D chemical structures are easier to obtain than those of proteins and that full\=/atom representations of ligands are not memory intensive.

Other data can also be incorporated to augment ground\=/truth information about PLIs. For example, molecular weights, polarity, and bioactive properties can be incorporated into models to further improve the prediction of PLIs~\cite{Aly_Abdelkader2023-tf, Li2022-pc}. Studies that included molecular weights, ligand polar surface area, and protein aromaticity~\cite{Aly_Abdelkader2023-tf}, or bioactive properties of chemical and clinical relevance~\cite{Li2022-pc} have resulted in improved predictions of binding affinity. Leveraging multiple\=/sequence alignment or phylogenetic information to identify co\=/evolutionary trends among AAs and sites of covalent modification has been shown to dramatically improve the accuracy of structural predictions of protein\=/ligand complexes~\cite{Abramson2024-qh}. The use of non\=/sequence/non\=/structural data can enable models to yield better predictive performance for characterizing protein and ligand and their interactions than models that do not~\cite{Aly_Abdelkader2023-tf}.

Given a protein\=/ligand representation, several predictive tasks are possible. \textit{Classification} studies seek to categorize PLIs into distinct groups, for example, whether a protein\=/ligand pair binds or not. These models are relatively simple and allow for input from various sources. \textit{Regression} studies use a continuous functional metric to characterize PLIs such as a binding affinity/dissociation constant ($K_{d}$) or inhibition constant ($IC50$). Continuous target variables allow for the involvement of numerical values derived directly from 'ground\=/truth' experimental data in both training and evaluation. Databases like PDBBind~\cite{Wang2004-tw} contain functional metrics such as $K_{d}$, and $IC50$, but not all protein and ligand pairings cataloged have such metrics available, for example, complexes identified from X\=/ray crystallography, Cryo-EM, or NMR screening studies~\cite{Vajda2006-vb, Chen2024-gg}. Since regression studies require quantitative PLI data and not merely whether a protein and ligand interact, relevant dataset sizes may be smaller than those for classification. However, gathering such data is a laborious process in terms of both time and laboratory resources.

Data for the study of PLIs can be manually curated by domain experts or sourced from existing datasets, such as PDBBind ~\cite{Wang2004-tw} and the Directory of Useful Decoys\=/Enhanced (DUD\=/E)~\cite{Huang2006-id, Mysinger2012-gp}, which includes tens of thousands of diverse pairings. Other datasets such as the Davis~\cite{Davis2011-hf} and KIBA~\cite{Tang2014-fz} datasets of kinase inhibitors, focus on particular types of proteins. While pre\=/assembled datasets are tempting to use out of convenience, relevant data need to be selected with an intended predictive task in mind. Table \ref{table:1} contains a collection of existing PLI datasets and databases for consideration.

\begin{table}[!ht]
\footnotesize
\centering
\scriptsize
    \rowcolors{1}{}{lightgray}
    \caption{Datasets and Databases for PLI Prediction}
    \begin{tabularx}{\textwidth}{>{\raggedright\arraybackslash}Xllllllr}
    \toprule
     \thead{\textbf{Dataset Name}}      & \thead{\textbf{Year}}       & \thead{\textbf{Proteins}}      & \thead{\textbf{Ligands}}
     & \thead{\textbf{Inter-} \\ \textbf{actions}}      & \thead{\textbf{Protein} \\ \textbf{Category}}      & \thead{\textbf{Ligand} \\ \textbf{Category}} \\
    \midrule
    \hline
    \underline{\textit{Functional Data Available}} &&&&&& \\
        Protein Data Bank (PDB) \cite{Berman2000-qn} & 2000 & 220,777 & - & - & General & General \\
        brenda~\cite{Schomburg2002-ax} & 2002 & 8,423 & 38,623 & - & Enzyme & General \\ 
        Natural Ligand Database (NLDB)\textsuperscript{+}~\cite{Puvanendrampillai2003-da} & 2016 & 3,248 & - & 189,642 & General & General \\ 
         DrugBank\textsuperscript{+}~\cite{Wishart2006-kc} & 2006 & 4,944 & 16,568 & 19,441 & \makecell[l]{Human \\ Proteome  } & General \\
        PDBBind\textsuperscript{+}~\cite{Wang2004-tw} & 2004 & - & - & 23,496 & General & General \\
        BindingDB~\cite{Liu2007-hq} & 2007 & 2294 & 505,009& 1,059,214 & General & General \\ 
         Davis-DTA~\cite{Davis2011-hf} & 2011 & 442 & 68 & 30,056 & Kinases & \makecell[l]{Kinase \\ Inhibitors} \\
        ChEMBL~\cite{Gaulton2012-jk} & 2012 & 15,398 & 2,399,743 & 20,334,684 & General & General \\ 
        DUD-E~\cite{Mysinger2012-gp} & 2012 & 102 & 22,886 & 2,334,372 & General & General \\    
        PSCDB~\cite{Amemiya2012-hr} & 2011 & - & - & 839 & \makecell*[l]{Human \\ Proteome} & General \\
        Iridium Database\textsuperscript{+}~\cite{Warren2012-ef} & 2012 & - & - & 233 & General & General \\
        KIBA~\cite{Tang2014-fz} & 2014 & 467 & 52,498 & 246,088 & Kinases & \makecell[l]{Kinase \\ Inhibitors} \\
        dbHDPLS\textsuperscript{+}~\cite{Zhu2019-lc} & 2019 & - & - & 8,833 & General & General \\        
         PDID~\cite{Wang2016-xz} & 2016 & 3,746 & 51 & 1,088,789 & \makecell*[l]{Human \\ Proteome} & General \\ 
        BindingMOAD~\cite{Smith2019-at} & 2020 & 11,058 & 20,387& 41,409 & General & General \\
        CovPDB\textsuperscript{+}~\cite{Gao2022-sp} & 2022 & 733 & 1,501 & 2,294 & General & General \\
        PSnpBind\textsuperscript{+}~\cite{Ammar2022-ck} & 2022 & 731 & 32,261 & 640,074 & General & General \\ 
        PLAS-5k\textsuperscript{+}~\cite{Korlepara2022-lf} & 2022 & - & - & 5,000 & Enzyme & - \\
        \makecell[l]{Protein\=/Ligand Binding \\ Database (PLDB)\textsuperscript{+}~\cite{Linge2023-lz}} & 2023 & 12 & 556 & 1831 & \makecell[l]{Carbonic \\ Anhydrase,\\ Heat Shock\\ Protein} & General \\ 
        BioLiP2~\cite{Wei2023-jh} & 2023 & 426,209 & - & 823,510 & General & General \\
        \underline{\textit{Functional Data Unavailable}} &&&&&&\\
        \makecell[l]{Database of Inter-\\acting Proteins~\cite{Xenarios2000-xs}} & 2004 & 28,850 & - & 81,923 & \makecell[l]{Different \\Species} & - \\
        \makecell[l]{Protein Small\=/Molecule\\ Database\textsuperscript{+}~\cite{Wallach2009-ka}} & 2009 & 4,916 & 8,690 & - & General & General \\
        CavitySpace\textsuperscript{+}~\cite{Wang2022-ky} & 2022 & 23,391 & - & 23,391 & General & General \\
    \bottomrule
    \end{tabularx}
    \begin{tablenotes}
      \footnotesize
      \item \textit{Note}: Published datasets may provide periodic updates in the future. Datasets marked with the “functional data available” label contain continuous metrics.
      \item -: Exact information is either not included in the source or is not readily obtainable.
      \item \textsuperscript{+}: Protein\=/ligand complexes are available with the dataset.
    \end{tablenotes}
    \label{table:1}
    \end{table}

\section{5. Machine Learning and NLP for PLIs}
\begin{figure}[t]
   \centering      
   \includegraphics[scale=.7]{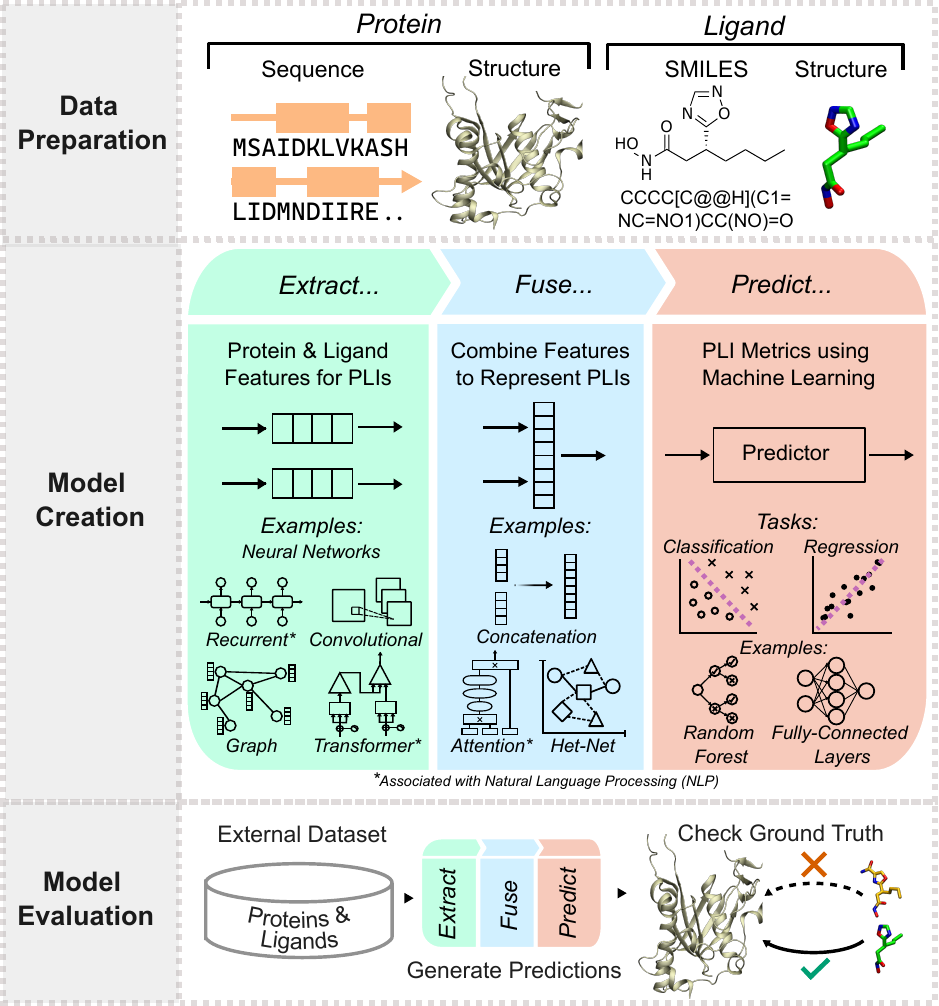}     
 \caption{Summary of the Data Preparation, Model Creation, and Model Evaluation Workflow. Model Creation for PLI studies follows an Extract\=/Fuse\=/Predict Framework: input protein and ligand data are extracted and embedded, combined, and passed into a machine learning model to generate predictions.}
 \label{fig:2}
\end{figure}

The general workflow for any ML\=/based study can be broadly characterized into three stages: data preparation, model creation, and model evaluation. A visual aid summarizing these stages is presented in Figure~\ref{fig:2}. For PLI studies, data preparation typically entails selecting the types and formats of protein and ligand data (e.g., sequence and/or structural). ML model creation may involve the following three tasks, although the boundary between these tasks could be fuzzy at times: (i) \textit{Extract}: the "extraction" of vector "embeddings" from the protein and ligand input data, which can be used in computational operations (described in Section 5.2) (ii) \textit{Fuse}: the fusion of protein and ligand vector embeddings, and (iii) \textit{Predict}: the prediction of a PLI target property as a model's output. The predictive capability of the model would be ideally validated against results from other studies and/or real\=/world measurements in a model evaluation stage. While data preparation and extraction steps have typically been the focus of most research efforts, every component of the workflow is crucial to successful PLI predictions.

\subsection{5.1. The Extract\=/Fuse\=/Predict Framework}
A variety of models for PLI prediction have been constructed in recent years, and these models tend to fall into four general categories: (1) \textit{sequence\=/based}, where protein sequences and SMILES are used to represent protein and ligand, respectively; (2) \textit{structure\=/based}, where structural information is included in the representation of both protein and ligand; (3) \textit{mixed representations}, where both structural and sequence information are involved; and (4) \textit{sequence\=/structure\=/plus}, which substantially incorporates other ground\=/truth information beyond sequence and structural data (such as molecular weights or polar surface area~\cite{Aly_Abdelkader2023-tf}). Tables 2 through 5 summarize several representative NLP\=/based PLI prediction studies across these categories over the past five years. Although PLI studies could be categorized in other ways—for example by the ML model used (neural network, decision tree, etc.) or by the predictive task type (classification vs. regression)—we have chosen to emphasize a categorization based on input data type since the computational methods used for sequence text and structural data comprise a major difference. 

\subsection{5.2. Extraction of Embeddings}
NLP approaches deconstruct text into individual tokens or "units of meaning” for use in computational operations and inferences via a process referred to as "tokenization". Schema for tokenization, aside from character\=/based and word\=/based, can also be sub\=/word\=/based. Sub\=/word\=/based tokenization breaks down text into units smaller than words to create a wider vocabulary; it is commonly selected when the definition of a "word" is unclear, as sub\=/words can be used as a means to discover "words"~\cite{Wang2019-qx, Liang2014-zt}. Common ways to assemble sub\=/words include methods such as “n\=/grams”, where each sub\=/word has a select fixed\=/length value $n$ (e.g., “Sma”, “mar”, “art”, etc. for $n$=3). While sub\=/word tokenization has been attempted in PLI studies for both protein (e.g., amino-acid k-mers such as "KHR", "LKL", "KGY" ) and ligand (e.g., "CCCC","[C@@H]")~\cite{Abbasi2020-zd, Zhou2023-ki, Zhou2021-ni, Ozcelik2021-hh, Gaspar2021-yl}, the current trend is to use amino acids and/or individual atoms directly as tokens.

To be processed computationally, tokens must be translated into a numerical form through a process known as “embedding”. There are many types of token embedding, but generally speaking, they are designed to capture either a particular token meaning, frequency, or both~\cite{Arseniev-Koehler2022-sr, Lake2023-dy} and represented by a multidimensional vector.  The direction of a token's vector embedding effectively represents its “meaning” while its magnitude represents the strength by which that meaning is conveyed. In isolation, each token could possess multiple meanings (e.g., the word "run" has multiple meanings~\cite{Winchester2011-yv}), and so context may be necessary to impart an intended meaning. NLP has been demonstrated to be highly effective at extracting patterns that convey context\=/dependent meanings from a large corpus of text. Embeddings that capture semantic meaning and relationships can then be used for many other tasks aside from predicting whether a protein interacts with a ligand, such as predicting protein and ligand solubilities~\cite{Panapitiya2022-gh, Wu2021-iz}. 

Token embedding is typically accomplished using a neural network (NN) architecture that approximates nonlinear relationships between the “inputs” of the network (the data) and its “outputs” (the predictions)~\cite{Krogh2008-fe}. Neurons in an artificial NN receive, integrate, and transmit signals to other neurons through a nonlinear response function and are arranged in layers. Information is passed from an input layer through one or more intermediate "hidden" layers to an output layer. Interconnection weights that govern the strength of influence of one neuron on another are crucial parameters of an NN. A wide variety of NNs have been applied to studying PLIs although not all are commonly used in NLP. Nevertheless, two types of NNs are commonly associated with NLP: Recurrent Neural Networks (RNNs) and attention\=/based NN models~\cite{Vaswani2017-gn}. Below, we highlight the details necessary to understand how RNNs, attention, and other non\=/NLP\=/driven NNs have been used to glean global patterns essential for PLI predictive tasks.

\subsubsection{5.2.1. Recurrent Neural Networks}
RNNs are specialized for processing sequential data in which the order of the data matters. Consider an input data sequence in which individual tokens are ordered by a time-step and embody a particular yet unknown pattern over the length of the sequence. In traditional NNs, information flows from the input layer to the output in a single pass, making it difficult to decipher any interdependencies between earlier tokens and subsequent ones. To remedy this, the RNN architecture introduces recurrence units through which the processing of the input sequence at the current time-step will also update the "hidden states" that nonlinearly capture the information of all input tokens up to the current time-step. These hidden states are functionally equivalent to the hidden layers of traditional NNs but differ by updating \textit{recurrently}, where information is carried over from previous time-steps to the current time-step. Thus, the dependencies between tokens of the sequential inputs can be captured. For example, given a protein sequence for which each AA is a token, an RNN would process the sequence of AAs one at a time to create and maintain a mapping for the next AA in the sequence accounting for all input tokens seen so far. While effective in many NLP tasks, early RNNs commonly suffered diminishing returns with increasing text length. This was due to systematic and nondiscriminatory retention of information from \textit{all tokens}, including outlier tokens that contribute little informationally to the underlying pattern.
 
To minimize diminishing returns, RNNs were modified to Long Short\=/Term Memory (LSTM) models. The signature component of LSTMs is the "forget gate", which selectively inhibits information not concordant with previously learned patterns found from processing prior tokens~\cite{Hochreiter1997-mc}. For example, in the task of predicting secondary structures, an LSTM's forget gate can attenuate the contribution of AAs that do not correlate with any defined secondary structural element~\cite{Sonderby2014-wc, Guo2019-yj}. Bidirectional LSTMs (BiLSTMs) have also been developed to capture both preceding and subsequent tokens in a sequence string by applying an LSTM to text in both original and reverse order, and concatenating each of the resulting embeddings end\=/to\=/end~\cite{Graves2005-ku}. 

LSTMs and BiLSTMs are promising embedding approaches for predicting binding affinities of proteins and ligands~\cite{Thafar2022-ig, Wei2022-cn, Yuan2022-mv}. However, their effectiveness has been limited by the size of the dataset that the LSTM/BiLSTM architecture can efficiently process. Most successful applications of LSTM to date have been applied to only relatively small training datasets, on the order of a few thousand proteins and ligand pairs. This limitation mainly arises from the inherently non\=/parallel design, which makes training on large datasets slow and computationally expensive. Thus, NN architectures that leverage parallelization will be important to ensure reasonable training and prediction runtimes. 

\subsubsection{5.2.2. Attention\=/Based Architectures}
Protein lengths can vary dramatically, from Insulin with 51\=/AAs to "giant proteins" that can exceed 85,000 AAs~\cite{West-Roberts2023-dt}. To use large amounts of sequence data to effectively process and predict PLIs for which long\=/distance interactions may be impactful, several alternatives to RNN have been proposed. The “neural attention"—or simply “attention”—mechanism is an important recent breakthrough by which "attention weights" are dynamically calculated to quantify the relative contribution of different input tokens or elements to a predictive end goal. In many NN architectures, attention can also incorporate hidden states into the calculation, allowing a more sophisticated mechanism for capturing longer\=/range correlations in deeper layers. 

\begin{figure}[t]
    \centering
    \includegraphics[scale=.4]{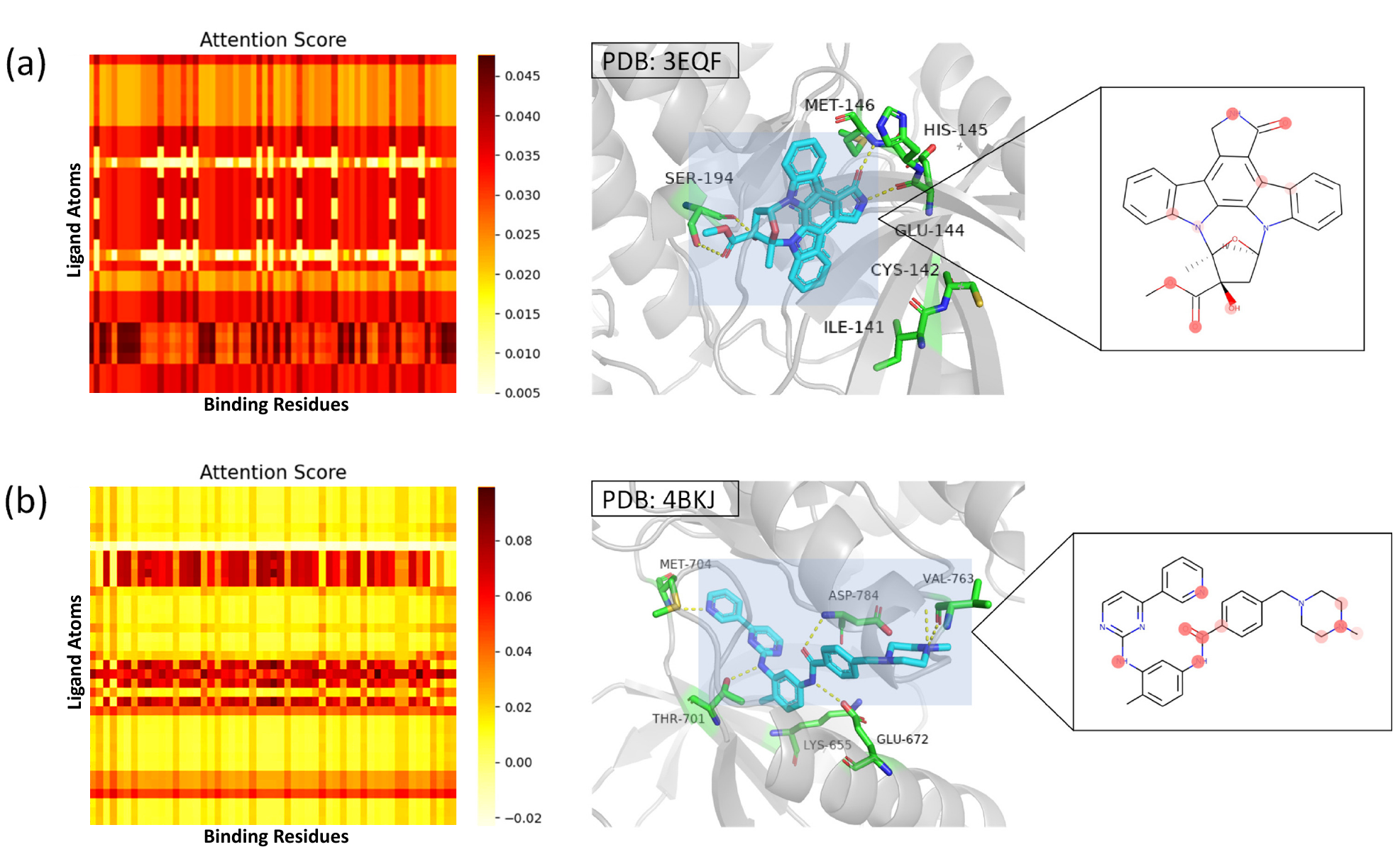}
    \caption{Sample Attention Weights for Relating Protein and Ligand. The heatmaps on the left help visualize the weighted importance of select protein residues and ligand atoms in a PLI. Structural views of the protein\=/ligand binding pocket are shown in the middle, with insets of the 2D ligand structures on the right. The colored residues and red color highlights indicate AAs in the protein binding pocket and ligand atoms with high attention scores. Adapted and modified from Figure 7 of Wu et al.\protect{\cite{Wu2024-ma}} used with permission under license CC BY 4.0.}                                           
    \label{fig:3}
\end{figure}

Attention mechanisms have proven highly compatible with traditional protein sequence analysis approaches in identifying long\=/distance interactions between AAs of a protein~\cite{Vig2020-ko}. In PLI studies, attention mechanisms can dynamically adjust the contribution of specific AAs or ligand atoms to a predictive outcome by amplifying interaction sites with higher attention scores and downplaying less relevant ones. This process mirrors the biological intuition that certain residues and atoms are more critical for binding in a protein\=/ligand complex than others. The use of attention mechanisms has enabled the identification of AAs in proteins and atoms in a ligand that are highly cross\=/correlated and appear to physically interact \cite{Koyama2020-li}, although the degree of success in identifying physically interacting sites remains to be carefully assessed. Attention has also provided an effective way to "fuse" protein and ligand representations in binding prediction models.~\cite{Chen2020-ns, Kurata2022-vd, Yuan2022-mv, Abbasi2020-zd, Zhao2021-do} Figure~\ref{fig:3} presents an example of how attention weights have been used to reveal potentially interacting sites between a protein and small molecule ligand. 

Attention is a versatile mechanism and can also be applied to structural information such as the spatial coordinates of individual atoms or contact maps of protein\=/ligand complexes.\cite{Jiang2020-nq, Nguyen2022-xu, Yu2023-ip}. The structural information of proteins and ligands can be well\=/represented by a graph with nodes representing AAs or atoms, and edges representing chemical bonds or amino acid contacts. Edges may also represent other predefined relationships or constraints between nodes.  Integrating attention mechanisms into Graph Neural Networks (GNNs), a class of NNs specialized for processing graphs, has been increasingly\=/used for the study of PLIs~\cite{Knutson2022-ub, Kyro2023-gs, Yousefi2023-qb, Yazdani-Jahromi2022-el}.  GNNs use "message\=/passing" whereby each node's embedding is updated iteratively based on information from connected nodes. Each connection can be assigned a weight that quantifies the likelihood of interdependence between connected nodes. For example, a cysteine residue may have a higher weight for a nearby cysteine than a nearby glycine due to the potential to form a disulfide bond between cysteines. GNNs are often augmented further, for example, by the addition of an attention mechanism to prioritize connected nodes during message\=/passing~\cite{Lim2019-dp, Kyro2023-gs, Yazdani-Jahromi2022-el, Yousefi2023-qb, Jiang2020-nq, Nguyen2022-xu}.

\subsubsection{5.2.3. Transformers}
While attention mechanisms have been quite beneficial for the predictive success of NLP methods, the "transformer" architecture pioneered in 2017 has also been instrumental in advancing these capabilities~\cite{Vaswani2017-gn}. Transformers are a type of NN architecture that divides attention mechanisms into multiple parallel operations, each applying a different set of weights to the input data sequence. Several relationships between tokens are captured and processed simultaneously, dramatically improving the efficiency with which human text can be processed. The transformer architecture is the foundation of popular large language models such as ChatGPT~\cite{Radford_undated-we} and is a key component of DeepMind’s AlphaFold system~\cite{Lin2023-jd, Jumper2021-xa}. Transformers have become widely used in bioinformatics, for DNA, RNA, and protein sequence analysis, as well as gene\=/based disease predictions and PLI predictions~\cite{Zhang2023-nb}.

Transformers are designed to solve the problem of “sequence transduction” or the conversion of an input sequence of ordinal data into a predicted output sequence, such as a translated text or a vector representation~\cite{Bahdanau2014-pq}. In NLP, this is called machine translation, whereby the input sequence, for example, could be a sentence in English and the output sequence is its French counterpart. Formally, the transformer architecture is an extension of the so\=/called "encoder\=/decoder" architecture, a state\=/of\=/the\=/art sequence\=/transduction method~\cite{Vaswani2017-gn}. The premise of  encoder\=/decoders is that sequentially ordered input data (e.g., English text, protein sequences, SMILES) can be “compressed” or encoded by a lower\=/dimensional fixed\=/length vector with minimal information loss. If the most "useful" or informative features of the data can be extracted and represented by this compressed or reduced vector representation, then the implicit rules/structures contained within the input data have been "encoded". Typically, in this reduced representation (called the “latent" space), inputs with similarly informative characteristics appear close to one another. These compressed vectors can subsequently be "decoded" or expanded to an output representation of choice to complete the transduction task. These transduction tasks naturally align with the goal of text translation from one language to another. Importantly, transformers differ from traditional encoder\=/decoder models by incorporating the attention mechanism. Attention allows latent representations to vary in length, thus eliminating a fundamental constraint of encoder\=/decoder models: that every input sequence, regardless of length, be represented by a fixed\=/length vector in the latent space. Given their improved precision, transformers are widely used today (especially for long input sequences), especially given their inherent parallel architecture, which makes processing datasets with billions of items feasible. As compared to LSTMs, transformers are architecturally more complex and tend to achieve better performance~\cite{Zeyer2019-rh, Irie2019-gj, Zouitni2023-ed, Parisotto2020-am}. Even so, transformer performance is not always the best, particularly when dealing with small datasets on the order of thousands of items~\cite{Bilokon2023-wf, Merity2019-je, Ezen-Can2020-hn}. In the biological domain, transformers have been applied to the prediction of protein\=/protein binding affinities~\cite{Unsal2022-hc}, post\=/translational modifications~\cite{Brandes2022-as}, and quantum chemical properties of small molecules~\cite{Luo2022-al}.

Early applications of transformers for the study of PLIs involved simply retraining existing models designed for human language inputs~\cite{Devlin2018-ww, Clark2020-da}; surprisingly,  these transformers surpassed existing state\=/of\=/the\=/art models for predicting binding affinities~\cite{Wang2022-an, Shin2019-yo}. As new transformers were developed specifically to handle protein sequence data, predictive performance for PLIs improved~\cite{Huang2020-ui, Shen2023-zl, Wang2022-eb}. These developments included preemptively dividing the texts into subsequences to determine which amino acids contribute to binding and merging embeddings from different transformers to provide multiple representational perspectives. More recently, transformers have been adapted for use with other data types, such as protein structures and images, as well as for predicting PLI properties beyond binding affinity, e.g., binding poses~\cite{Qian2022-su,Zhou2023-ki}.

\subsection{5.3. Fusion of Protein\=/Ligand Representations: Concatenation or Cross-Attention}
Once candidate interacting protein and ligand embeddings are extracted, they need to be fused for an interaction pattern to emerge. Method development for extracting embeddings from protein and ligand sequence data has been the primary focus to date such that approaches for fusion have been somewhat neglected, although they have garnered greater interest in recent years. A naive method for fusion is to simply end\=/to\=/end concatenate protein and ligand embedding vectors. A more refined approach could involve advanced data structures like graphs, whereby information such as coordinates of protein and ligand is used not only to build a graph representation but is also incorporated into an attention mechanism to account for local factors such as polarity or size~\cite{Nguyen2020-eq, Kyro2023-gs, Yu2023-ip}. A mechanism of “cross\=/attention” could be incorporated into the fusion approach whereby the importance between the different \textit{token representations} of the protein and ligand are directly calculated ~\cite{Zhao2021-do, Kurata2022-vd, Jin2023-qf} in an attempt to mirror the underlying interaction of a protein with a ligand~\cite{Knutson2022-ub}. Cross\=/attention has been shown to be at least as competitive in predictive PLI tasks as other fusion methods~\cite{Nguyen2020-eq}, and an improvement over the use of separate, independent attention mechanisms for both protein and ligand~\cite{Nam2017-gz}. 

While fusion appears to be a natural and important component for NLP studies of PLIs, some models circumvent the idea of fusion altogether in lieu of  considering only protein or ligand representations alone. For example, Wang et al.’s CSConv2D algorithm only embeds ligand information~\cite{Wang2021-um}. An individual model is trained separately for each protein to predict that protein’s compatible ligands, resulting in the creation of hundreds of models. Although the study focuses on predicting PLIs, protein information is only incorporated indirectly by labeling ligands as either binding or non\=/binding to a given protein during its model's training. Nonetheless, protein\=/only or ligand\=/only models are rare, with most contemporary NLP\=/PLI models considering both protein and ligand together through a fusion step. 

Mixed\=/data approaches aimed at combining different data types for protein and/or ligand (e.g., sequence + structure; sequence + image~\cite{Qian2022-su}; or both sequence and structure for protein + structure for ligand~\cite{Jin2023-qf}) have further spurred studies into which input format is best for the protein or ligand. Mixed\=/data models may use a variety of architectures such as an LSTM or transformer for a protein sequence and a GNN for ligand structures~\cite{Mukherjee2022-xg, Chen2020-ns}. Combining multiple state\=/of\=/the\=/art embeddings for both sequence and structure has outperformed sequence\=/only baselines~\cite{Chen2020-ns}. Despite the increased complexity involved in handling sequence and structural data simultaneously, mixed\=/data models are advantageous for both the ease\=/of\=/use of protein sequences and the completeness of ligand structural representations.

Although under\=/explored, combining multiple embeddings for each protein and ligand input in the fusion process may be beneficial. It has been suggested that different protein encoders for extracting features may gather different but relevant information to improve predictive outcomes~\cite{Anteghini2023-lr}. In the algorithm, DeepPurpose, Huang et al. pursued a library approach that offered fifteen different protein and ligand embeddings (including transformer and RNN) to be combined and fed into a small NN to generate binary binding and/or continuous binding affinity predictions~\cite{Huang2020-fb}. This menu-option approach enables users to compare feature extractors and find the best protein and ligand embeddings for their research. Another approach is to combine multiple embeddings through operations such as component-wise multiplication or component-wise difference, as each embedding could represent a different set of features~\cite{Shen2023-zl,Anteghini2023-lr}. Shen et al.'s SVSBI algorithm~\cite{Shen2023-zl}  demonstrated how a higher\=/order embedding, by concatenating three different transformer embeddings, could outperform several state\=/of\=/the\=/art baselines in the prediction of binding affinity, including those based on individual transformers alone.

\subsection{5.4 Prediction of Target Variables}
Ultimately, specific research questions must motivate which relevant PLI target variables are to be predicted by the ML models constructed. These models often consist of one or more fully\=/connected layers with relatively few parameters than the NNs used for feature extraction or fusion. The purpose of these layers is to utilize the latent protein and ligand features to predict an output target variable such as binding affinity or a binary indication of whether a pairing interacts. Thus, the fused protein and ligand embeddings are passed through these final layers to compute the prediction. Embeddings that effectively capture important underlying features can also be applied to predict other useful properties beyond binding affinity such as protein and ligand solubility~\cite{Panapitiya2022-gh, Wu2021-iz}. 

\begin{table}[!ht]
\caption{Sequence-Based PLI Prediction Models}
\scriptsize
\centering
\begin{threeparttable}
\begin{tabularx}{\textwidth}{p{2.5cm} p{3cm} p{3cm} p{2.9cm} p{3cm}} 
\toprule
{\multirow{2}{*}{\textbf{Model Name}}} & \multicolumn{2}{c}{\makecell{\textbf{Extraction}}} & {\multirow{2}{*}{\makecell{\textbf{Fusion}}}} & {\multirow{2}{*}{\makecell{\textbf{Prediction}}}}\\ 
\cline{2-3}
\addlinespace[1mm]
& {\textbf{Protein Extractor}} & {\textbf{Ligand Extractor}} & & \\
\hline
\bottomrule
\rowcolor{lightgray}
\multicolumn{1}{>{\columncolor[gray]{.8}}l}
{\underline{\textit{LSTM}}} &&&&\\
    Affinity2Vec\cite{Thafar2022-ig} & ProtVec & Seq2Seq & Heterogeneous \newline Network & Gradient-Boosting Trees (R) \\
    \rowcolor{lightgray}
    DeepLPI~\cite{Wei2022-cn} & ResNet & ResNet & Concatenation with LSTM & FCN (C, R) \\
    FusionDTA~\cite{Yuan2022-mv} & BiLSTM & BiLSTM & Concatenation with Linear Attention & FCN (R) \\
\rowcolor{lightgray}
\multicolumn{1}{>{\columncolor[gray]{.8}}l}
{\underline{\textit{Transformer}}} &&&&\\
    Shin et al.~\cite{Shin2019-yo} & CNN & Transformer & Concatenation & FCN (R) \\
    \rowcolor{lightgray}
    MolTrans~\cite{Huang2020-ui} & Transformer  & Transformer & Interaction Matrix$^a$ with CNN & FCN (C) \\
    ELECTRA-DTA~\cite{Wang2022-an} & CNN with Squeeze-and-Excite Mechanism & CNN with Squeeze-and-Excite Mechanism  & Concatenation & FCN (R)\\
    \rowcolor{lightgray}
    MGPLI~\cite{Wang2022-eb} & Transformer, CNN & Transformer, CNN& Concatenation & FCN (C) \\
    SVSBI~\cite{Shen2023-zl} & Transformer, LSTM, and AutoEncoder  & Transformer, LSTM, and AutoEncoder & k-embedding fusion$^b$ & FCN, Gradient-Boosting Trees$^c$ (R) \\
\rowcolor{lightgray}
\multicolumn{2}{>{\columncolor[gray]{.8}}l}{\underline{\textit{Non-Transformer Attention}}} &&&\\
    DeepCDA~\cite{Abbasi2020-zd} & CNN with LSTM & CNN with LSTM & Two-Sided \newline Attention$^c$ & FCN (R)\\
    \rowcolor{lightgray}
    HyperAttention- DTI~\cite{Zhao2021-do} & CNN & CNN & Cross-Attention,\newline Concatenation & FCN (C)\\
    ICAN~\cite{Kurata2022-vd} & Various & Various & Cross-Attention,\newline Concatenation & 1D CNN (C) \\
\rowcolor{lightgray}
\multicolumn{2}{>{\columncolor[gray]{.8}}l}
{\underline{\textit{Other NLP Methods}}} &&&\\
    GANsDTA~\cite{Zhao2019-zg} & GAN Discriminator & GAN Discriminator & Concatenation & 1D CNN (R) \\
    \rowcolor{lightgray}
    Multi-PLI~\cite{Hu2021-iq} & CNN & CNN & Concatenation & FCN (C, R) \\
    ChemBoost~\cite{Ozcelik2021-hh} & Various & SMILESVec & Concatenation & Gradient-Boosting Trees (R) \\ 
\bottomrule
\end{tabularx}
\begin{tablenotes}
      \scriptsize
      \item \textit{Note:} A model's task of Classification (C) and/or Regression (R) is denoted beside the "Prediction" column entries in parenthesis. Definitions for specific terms may be found in the Glossary (Table \ref{Glossary}).
      \item \underline{\textit{Terms Defined by the Cited Authors}}: $^a$\textbf{Interaction Matrix}: Output from dot product operations to measure interactions between protein sub\=/sequence and ligand sub\=/structure pairs. $^b$\textbf{k\=/embedding fusion}: The use of machine learning to find an optimal combination of lower\=/order embeddings via different integrating operations. $^c$\textbf{Two\=/sided Attention}: Attention mechanism that computes scores using the products of both pairs of protein/ligand fragments and protein/ligand feature vectors.
      \end{tablenotes}
\label{table:2}
\end{threeparttable}
\end{table}

\begin{table}[!ht]
\caption{Structure-Based PLI Prediction Models}
\scriptsize
\centering
\begin{threeparttable}
\begin{tabularx}{\textwidth}{P{2.5cm} P{3cm} P{3cm} P{2.9cm} P{3cm}} 
\toprule
{\multirow{2}{*}{\textbf{Model Name}}} & \multicolumn{2}{c}{\textbf{Extraction}} & {\multirow{2}{*}{\textbf{Fusion}}} & {\multirow{2}{*}{\textbf{Prediction}}}\\ 
\cline{2-3}
\addlinespace[1mm]
& {\textbf{Protein Extractor}} & {\textbf{Ligand Extractor}} & & \\
\hline
\bottomrule
\rowcolor{lightgray}
\underline{\textit{Transformer}} &&&&\\
    UniMol~\cite{Zhou2023-ki} & Transformer-Based Encoder & Transformer-Based Encoder & Concatenation & Transformer-Based Decoder (R)\\
\rowcolor{lightgray}
\underline{\textit{Other Attention}} &&&&\\
    Lim et al.~\cite{Lim2019-dp} & GNN & GNN  & Attention & FCN (C) \\
    \rowcolor{lightgray}
    Jiang et al.~\cite{Jiang2020-nq} & GCN & GCN & Concatenation & FCN (R) \\
    GEFA~\cite{Nguyen2022-xu} & GCN & GCN & Concatenation & FCN (R)\\
    \rowcolor{lightgray}
    Knutson et al.~\cite{Knutson2022-ub} & GAT & GAT & Concatenation & FCN (C, R) \\
    AttentionSite-DTI~\cite{Yazdani-Jahromi2022-el} & GCN with Attention & GCN with Attention & Concatenation, \newline Self-Attention & FCN (C, R) \\
    \rowcolor{lightgray}
    HAC-Net~\cite{Kyro2023-gs} & GCN with Attention \newline Aggregation & GCN with Attention & Combined Graph \newline Representation & FCN (R) \\
    BindingSite-AugmentedDTI~\cite{Yousefi2023-qb} & GCN with Attention & GCN with Attention & Concatenation, \newline Self-Attention & Various (R)\\
    \rowcolor{lightgray}
    PBCNet~\cite{Yu2023-ip} & GCN & Message-Passing NN  & Attention & FCN (R)\\
\bottomrule
\end{tabularx}
\begin{tablenotes}
      \scriptsize
      \item \textit{Note}: A model's task of Classification (C) and/or Regression (R) is denoted beside the "Prediction" column entries in parenthesis. Definitions for specific terms may be found in the Glossary (Table \ref{Glossary}).
      \end{tablenotes}
\label{table:3}
\end{threeparttable}
\end{table}

\begin{table}[!ht]
\caption{Mixed Representation PLI Prediction Models}
\scriptsize
\centering
\begin{threeparttable}
\begin{tabularx}{\textwidth}{P{2.2cm} P{1.5cm} P{2.7cm} P{2.7cm} P{2.4cm} P{2.5cm}}
\toprule
\multirow{1}{*}{\textbf{Model}} & \multirow{1}{*}{\textbf{Input}} & \multicolumn{2}{c}{\textbf{Extraction}} & \multirow{3}{*}{\textbf{Fusion}} & \multirow{3}{*}{\textbf{Prediction}}  \\ 
\cline{3-4}
\addlinespace[1mm]
{\textbf{Name}} & {\textbf{Type}} & {\textbf{Protein}} & {\textbf{Ligand}} & & \\
\hline
\bottomrule
\rowcolor{lightgray}
\underline{\textit{LSTM}} &&&&&\\
    Zheng et al.~\cite{Zheng2019-gq} & P: Struct. \newline L: Seq. & Dynamic CNN$^a$ with Attention & BiLSTM with Attention & Concatenation & FCN (C) \\ 
    \rowcolor{lightgray}
    DeepGLSTM~\cite{Mukherjee2022-xg} & P: Seq. \newline L: Struct. & BiLSTM with FCN & GCN & Concatenation & FCN (R)  \\
\underline{\textit{Transformer}} &&&&&\\
    \rowcolor{lightgray}
    Transformer-
    \newline CPI~\cite{Chen2020-ns} & P: Seq. \newline L: Struct. & Transformer Encoder & GCN & Transformer Decoder & FCN (C) \\ 
    DeepPurpose~\cite{Huang2020-fb} & P: Seq. \newline L: Either & 4 Various Encoders & 5 Various Encoders & Concatenation & FCN (C, R) \\
    \rowcolor{lightgray}
    CAT-CPI~\cite{Qian2022-su} & P: Seq. \newline L: Image & Transformer Encoder & Transformer Encoder & Concatenation & CNN and FCN (C) \\
\underline{\textit{Non-Transformer Attention}} &&&&&\\
    \rowcolor{lightgray}
    Tsubaki et al.~\cite{Tsubaki2018-yt} & P: Seq. \newline L: Struct.  & CNN & GNN & Attention and Concatenation & FCN (C) \\
    DeepAffinity~\cite{Karimi2019-ql} & P: Seq. \newline L: Struct. & RNN-CNN with Attention & RNN-CNN with Attention & Concatenation & FCN (R) \\
    \rowcolor{lightgray}
    MONN~\cite{Li2020-jl} & P: Seq. \newline L: Struct. & CNN & GCN &  Pairwise Interaction Matrix$^b$, Attention & Linear Regression (C, R) \\
    GraphDTA~\cite{Nguyen2020-eq} & P: Seq. \newline L: Struct. & CNN & 4 GNN Variants & Concatenation & FCN (R)\\
    \rowcolor{lightgray}
    CPGL~\cite{Zhao2023-eb} & P: Seq. \newline L: Struct. & LSTM & GAT with Attention & Two-Sided Attention$^c$,\newline Concatenation & Logistic Regression (C) \\
    CAPLA~\cite{Jin2023-qf} & P: Both \newline L: Struct. & Dilated Convolutional Block & Dilated Convolutional Block with Cross-Attention to Binding Pocket & Cross-Attention, \newline Concatenation & FCN (R)\\
\bottomrule
\end{tabularx}
\begin{tablenotes}
      \scriptsize
      \item \textit{Note}: A model's task of Classification (C) and/or Regression (R) is denoted beside the "Prediction" column entries in parenthesis. Definitions for specific terms may be found in the Glossary (Table \ref{Glossary}). The input representations for sequence and structure are abbreviated for brevity.
      \item \underline{\textit{Terms Defined by the Cited Authors}}: $^a$\textbf{Dynamic CNN}: ResNet\=/based CNN modified to handle inputs of variable lengths by padding the sides of the input with zeroes. $^b$\textbf{Pairwise Interaction Matrix}: A [number of atoms]-by-[number of residues] matrix in which each element is a binary value indicating if the corresponding atom\=/residue pair has an interaction.~\cite{Li2020-jl} $^c$\textbf{Two\=/sided Attention}: Attention mechanism that uses dot product operations between protein AA and ligand atom pairs, while taking matrices of learned weights into account.
      \end{tablenotes}
\label{table:4}
\end{threeparttable}
\end{table}

\begin{table}[!ht]
\caption{Sequence-Structure-Plus PLI Prediction Models}
\scriptsize
\centering
\begin{threeparttable} 
\begin{tabularx}{\textwidth}{P{2cm} P{2.6cm} P{2.6cm} P{2.6cm} P{2.2cm} P{2cm}}
\toprule
\multirow{3}{*}{\textbf{Model Name}} & \multicolumn{3}{c}{\textbf{Extraction}} & \multirow{3}{*}{\textbf{Fusion}} & \multirow{3}{*}{\textbf{Prediction}} \\ 
\cline{2-4}
\addlinespace[1mm]
& {\textbf{Protein \newline Extractor}} & {\textbf{Ligand \newline Extractor}} & {\textbf{Additional \newline Features Used}} & & \\
\hline
\bottomrule
\rowcolor{lightgray}
\multicolumn{2}{>{\columncolor[gray]{.8}}l}
{\underline{\textit{LSTM}}}&&&&\\
    HGDTI~\cite{Yu2022-nl} & BiLSTM & BiLSTM & Disease and Side Effect Information & Concatenation & FCN (C) \\
    \rowcolor{lightgray}
    ResBiGAAT~\cite{Aly_Abdelkader2023-tf} & Bidirectional GRU with \newline Attention & Bidirectional GRU with \newline Attention & Global Protein Features & Concatenation & FCN (R) \\
\underline{\textit{Transformer}} &&&&&\\
    \rowcolor{lightgray}
    Gaspar et al.~\cite{Gaspar2021-yl}& Transformer or LSTM & ECFC4 Fingerprints & Multiple Sequence \newline Alignment Information & Concatenation & Random Forest (C) \\
    HoTS~\cite{Lee2022-ji} & CNN & FCN & Binding Region & Transformer Block & FCN (C, R) \\
    \rowcolor{lightgray}
    PLA-MoRe~\cite{Li2022-pc} & Transformer & GIN and AutoEncoder & Bioactive Properties & Concatenation &  FCN (R) \\
    AlphaFold 3~\cite{Abramson2024-qh} & Attention-Based Encoder$^a$ & Attention-Based Encoder$^a$ & Post-Translational Modifications, Multiple Sequence Alignment Information & Attention & Diffusion Transformer$^b$\\ 
\rowcolor{lightgray}
\multicolumn{2}{>{\columncolor[gray]{.8}}l}
{\underline{\textit{Other NLP Methods}}} &&&&\\
    MultiDTI~\cite{Zhou2021-ni} & CNN with FCN & CNN with FCN & Disease and Side Effect Information & Heterogeneous Network & FCN (C) \\
\bottomrule
\end{tabularx}
\begin{tablenotes}
      \scriptsize
      \item \textit{Note}: A model's task of Classification (C) and/or Regression (R) is denoted beside the "Prediction" column entries in parenthesis. Definitions for specific terms may be found in the Glossary (Table \ref{Glossary}).
      \item \underline{\textit{Terms Defined by the Cited Authors}}: $^a$\textbf{*Atom Attention Encoder}: An attention\=/based encoder that uses cross\=/attention to capture local atom features. $^b$*\textbf{Diffusion Transformer}: A transformer\=/based model that aims to remove noise from predicted atomic coordinates until a suitable final structure is output.
      \end{tablenotes}
\label{table:5}
\end{threeparttable}
\end{table}

\subsection{5.5. Evaluation}
Evaluation is typically performed by comparing statistical metrics between models on the same test datasets. Evaluation metrics vary by task: classification predictions can be assessed via precision, recall, and F1 score metrics whereas regression predictions are often evaluated relative to the ground\=/truth test data via concordance index and mean square error metrics~\cite{Gonen2005-tm, M2006-wz}. Pre-made datasets such as PDBBind~\cite{Wang2004-tw} frequently come with both training and test datasets to enable fair comparisons with other established models. Models aiming to be generalizable across several types of PLIs should ideally be evaluated on several different sets of proteins and ligands. 

While ML models can be assessed through the aforementioned statistical metrics, the practical utility of PLI predictive models and their predictive accuracy in real\=/world cases is best validated by PLI domain experts in the field~\cite{Deller2015-oe}. For example, if a model is designed to predict binding affinities, a set of predictions generated \textit{in silico} would be best confirmed through \textit{in vitro} experimentation. This can serve two purposes: justifying a model's use where it can be most effective and creating an opportunity for future interdisciplinary collaboration between ML practitioners and PLI domain experts in computational and experimental biology.

\section{6. Challenges and Future Directions}
Generative AI and NLP techniques have revolutionized how we tackle tasks related to human language. Early successes of NLP methods in discerning the "rules" of protein structure (as exemplified by AlphaFold~\cite{Jumper2021-xa}) suggest significant potential for NLP to transform our approach to studying PLIs. While many innovations in the NLP computational toolkit for PLIs have emerged in recent years, several practical hurdles remain, limiting the impact and potential insights derivable from the ML approaches. This section presents an overview of the many challenges confronting the PLI field and suggests various avenues to address them. 

\subsection{6.1. Lack of "True Negatives"}
A common challenge in today's data\=/driven ML paradigm is the limited availability of abundant, high-quality, and labeled data. In PLI studies, there is a particular lack of bona fide “negative examples”, i.e., data for ligand\=/like molecules that do not bind a protein of interest that are critical for model training. In "supervised" ML~\cite{Nasteski2017-xs}, models are trained on data with labels of whether a protein-ligand pair is binding or non\=/binding, and protein\=/ligand data spanning the full spectrum of interaction/no\=/interaction are necessary for models to 'learn'. When a similar situation is encountered in other ML problems, a common approach is to select random data points not explicitly labeled as "positive" and to declare them as “negative”.  This would be equivalent to assigning random ligands to each protein and treating them as negative PLI examples. Unfortunately, given the complexity and specificity of PLIs, these are often \textit{trivial} negative examples since molecules that do not interact with a protein of interest \textit{and} are dissimilar to the "true" ligands embody little information from which ML models can learn. Manually curating protein\=/ligand pairs that display weak interactions or lower binding affinities is an option for addressing this problem, although this is time\=/consuming and labor\=/intensive. 

Acquiring the requisite negative data for classification studies is tied to experimental studies that conclusively determine whether pairings bind. The availability of informative negative PLI data requires deliberate efforts of domain experts who recognize the importance of generating, curating, and reporting such data, which are rarely publicized or emphasized in the literature regardless of data type~\cite{Kozlov2024-ce, Edfeldt2024-sl, Mlinaric2017-oi}. Learning from positive data only or with unlabeled data is, therefore, an active field of study, with many attempts applying"unsupervised" or "semi\=/supervised" methods~\cite{Albalate2013-wk} (see~\cite{Zhao2019-zg, Sajadi2021-ld} for examples related to PLI prediction). Compared with supervised models, un\=//semi\=/supervised models typically require larger datasets of tens to hundreds of thousands of PLIs. Furthermore, the associated network architectures may be more computationally intensive~\cite{Zhao2019-zg}. In cases where negative data does exist but at a significantly reduced quantity, additional remedies may be attempted. For example, classification studies of PLIs can adjust the distribution of ligands to ensure \textit{equal proportions} in the positive and negative examples represented; this has been shown to mitigate the issue of an overabundance of positive data~\cite{Najm2021-jr}. Future studies should resolve the lack of readily available non\=/interacting protein\=/ligand pairs, perhaps through mining the scientific literature for meaningful non\=/binding pairs.

\subsection{6.2. Diversity Bias in PLI Datasets}
Many PLI datasets possess underlying bias with respect to either the diversity or types of proteins and ligands represented, which hinders the effectiveness of ML algorithms. Training with \textit{insufficiently different} data points can lead to poor generalizability and predictive performance when extended to real-world examples not represented in the training dataset. For example, binding affinity predictors trained on the popular PDBBind dataset~\cite{Wang2004-tw} with both protein and ligand information represented performed no better than those trained on protein- or ligand\=/only information~\cite{Yang2020-ns}. This implies that the PDBBind dataset is biased and that the protein\=/ligand trained model failed to discern the mechanics of binding and rather "memorized" the most popular representatives or non\=/informative patterns within the dataset. The commonly used DUD\=/E~\cite{Mysinger2012-gp} dataset of bioactive compounds and respective protein targets was found to suffer from a similar problem: PLI binding classification models that differentiated binders/non\=/binders to a high degree of accuracy resulted only because the binders and non\=/binders were of different shape classes and not because they embedded any relevant information about the protein\=/ligand interface~\cite{Yang2020-ns, Sieg2019-nq}. The existing literature suggests that this is a problem of quality over quantity, as memorization\=/related biases in PLI models are \textit{not} alleviated by merely increasing the dataset size or removing over\=/represented items.~\cite{Volkov2022-vm} The presence of bias is understandable, given how idiosyncratic research interests in biological or pharmaceutical fields shape the particular proteins and subsets of ligands that are studied and the type of PLI data generated and made available. Given that models trained on biased data often fail in practical, real\=/world prediction tasks, the creation of high\=/quality, well\=/balanced, and unbiased PLI datasets is essential to the future of ML\=/based PLI studies and should be made a priority.

To train more generalizable models, systematic datasets with proteins and ligands beyond those of biological and pharmaceutical interest need to be evaluated. One way around the experimental challenges of generating sufficient protein\=/ligand data may be through high\=/throughput molecular dynamics simulations and/or docking studies of PLIs using AlphaFold\=/predicted \cite{Jumper2021-xa} protein structures. Although current simulation methods are time intensive, the availability of powerful computing clusters and trends towards increasingly powerful GPU hardware may make this approach feasible in the not\=/too\=/distant~future, and the benefits may be worth investing in this pursuit. This approach could be automated, requiring far less human intervention than laboratory experiments, and can yield valuable binding pocket information for better structure-based ML predictions.

\subsection{6.3. The Limitations of "Language-ness" in Protein and Ligand Text Representations}
As compared to human languages, both proteins and ligands have significant structural and ontological differences that have to be accounted for when designing a modelhe following nuances have driven investigation into modifying existing NLP architectures to accommodate for protein and ligand representations.

In linguistics, a “word” is a complete unit of meaning that a reader can recognize. However, for protein sequences, such corresponding units are not easily demarcated. It would be dubious to assume AAs are equivalent to “words” because the roles of individual AAs are highly dependent on their context and environment. The meaning of a word in a human language may be independent of its surroundings if the word has only one definition; however, an amino acid has "meaning" across several levels/dimensions, influencing secondary structure, tertiary structure, motif function, and/or binding interactions. Conversely, it is also difficult to view motifs or domains as “words” since not all regions of a protein are independent of one another~\cite{Takahashi1996-ze}. This lack of word\=/equivalence is one motivation behind "sub\=/word" tokenization methods that attempt to discern a hierarchy of word equivalents in protein sequences~\cite{Liang2014-zt}. Protein sequences are also different from human languages in the length scale of interactions and the number of long\=/distance interactions that contribute to a 3D structure. While human language texts have distant relationships, such as between the subject and pronoun of a sentence or a passage that foreshadows another in an essay, these relations can be deduced by a reader and remain relatively sparse on a per\=/sentence basis. In contrast, AAs have numerous distant relationships and cannot be thoroughly predicted by even an expert in protein biochemistry without the assistance of computational or experimental tools, thus adding a layer of complexity to the analysis of AA sequences that can be well compensated for by ML approaches.

In the chemical space, the SMILES format is dissimilar from human languages in the large variation in text length and in the difficulty of finding an ideal tokenization scheme. First, the lengths of SMILES strings could vary even more than those of protein sequences, ranging from listing each atom of small molecules to those constituting entire proteins, although protein\=/protein interactions are generally considered a separate problem. The SMILES format is less practical to use for larger molecules, since structural graphs can provide a more compact and accurate representation of atoms in a large three\=/dimensional structure. A further disadvantage of using SMILES is that it is difficult to intuitively discern “word” equivalents within the string. Individual branches separated by parenthesis could be viewed as words~\cite{Lee2022-vq}, but this is only practical for small branching groups. Moreover, the handling of nesting parentheses in SMILES ge molecules can be problematic and has become a major limiting factor in ML models designed to generate novel molecules~\cite{Skinnider2024-jc}. The sum of these SMILES shortcomings has led to the development of alternative chemical representations for computational studies such as DeepSMILES and SELFIES~\cite{OBoyle2018-vn, Krenn2020-yf}. Although promising, these alternate forms have rarely been used in PLI studies to date. The question remains whether a three\=/dimensional molecule can be truly mapped to a text representation in a way that preserves all relevant structural information for PLIs. 

\subsection{6.4. Interpretable Design in PLI Predictions}
Catalyzed by the open-data movement and widely accessible machine-learning tools, hidden or explicit patterns are discovered from datasets through weighted mathematical operations that are difficult to interpret and yet many effective predictive models have been developed. A majority of ML studies fail to consider designing human-friendly interpretations of \textit{how} their models' predictions are calculated. This is a significant contemporary challenge that has prompted the growth of explainable AI (XAI) as an active research field. 

To build trust among biologists and broaden scientific acceptance, future ML models must be more understandable to end-user biologists than provided by common “black\=/box” designs~\cite{Adadi2018-nl}. One potential approach for bridging the "explainability" gap in PLI studies is the use of attention weights to corroborate existing protein\=/ligand contacts (cf.~Fig.~\ref{fig:3})~\cite{Chen2020-ns, Kurata2022-vd, Yuan2022-mv, Abbasi2020-zd, Zhao2021-do}. Attention weights provide a degree of interpretability by highlighting binding regions in PLI models that converge with higher weight values. Given the reality of "false positives" whereby higher binding weights are inadvertently assigned to non-binding regions, attention weights alone may not constitute a satisfactory basis for explaining or inferring what regions govern binding interactions.  Unfortunately, a systematic assessment of ‘false positives’ in attention weights has yet to be performed, leaving it unclear whether they are a reliable metric\cite{M2006-wz}. Such potential inaccuracies are but one facet of a larger debate on whether attention weights provide sufficient explanatory power for PLI models\cite{Bibal2022-wl}. 

While NLP presents attention mechanisms as one possible avenue, other methods of explainability have also started to be applied to the study of PLIs. One example is the use of a game\=/theoretical approach to compute "Shapley values", which quantify the importance of individual features by evaluating each feature's contribution to the final prediction across all possible combinations of those features~\cite{Lundberg2017-dw, Gu2023-qw}. Visualization may also be a great tool for identifying possible binding interactions. For example, graph visualization can help depict the bonds between an interacting protein and ligand, and "saliency maps" can highlight specific subregions of protein and ligand that are the most influential in the model's prediction by determining how changes in individual input features affect the output~\cite{Rodis2023-eh}. Several underutilized avenues for establishing interpretability remain~\cite{Gilpin2018-nh}, but none have been established as a state-of-the-art; determining a standard method of interpretability for PLI prediction models will be critical for the field.

\subsection{6.5. The Insufficiency of an NLP\=/only Approach for PLI Studies?}
While NLP offers beneficial strategies for the study of PLIs, it is far from a panacea and is often complemented by insights from other disciplines. Existing attention\=/based, state\=/of\=/the\=/art NLP models are limited by the need for substantial amounts of training data for the best results~\cite{Popel2018-ki}. There may be opportunities for other disciplines of computer science to contribute positively to the PLI field. For example, it may be more fruitful to incorporate computer vision techniques that are better at handling structural information over NLP techniques that are designed for handling text \cite{Eguida2020-mo}. 

Several studies have combined NLP with more unusual architectures or complementary approaches. For example, Zhao et al. created an algorithm that uses so\=/called generative adversarial networks (GANs) as a means of embedding protein sequences and SMILES independently~\cite{Zhao2019-zg}. GANs feature a dual NN architecture: a generator that creates artificial data points and a discriminator that is trained to distinguish between real and artificial data. Both components were trained together in a process akin to an evolutionary arms-race, as the discriminator repeatedly learns to identify key features of the input that help distinguish real data from increasingly realistic artificial data. Zhao et al. demonstrated competitive results relative to selected benchmarks even though the efficacy of their GAN was stated to be limited by the small dataset used but would likely perform better if trained on at least thousands of diverse proteins and ligands \cite{Zhao2019-zg}. In other studies, computer vision methods have been used to combine images of proteins or ligands as inputs alongside their text representations; features across the two modalities enable attention mechanisms to capture cross\=/feature correlations across data types~\cite{Zheng2019-gq, Qian2022-su}. While the success of the aforementioned hybrid strategies did not exceed the performance of other neural networks in PLI predictive tasks~\cite{Ozturk2018-zc}, they demonstrate the potential for innovation using advances from other sub\=/domains of ML and computer science beyond NLP.

Biological domain knowledge is crucial both for framing the computational challenges related to ML and for identifying best practices for handling protein and ligand data. Approaches that are grounded in a deep understanding of the underlying domain\=/specific science have proven to be forerunners in the practical success of ML methods, as demonstrated by the AlphaFold initiative's sophisticated use of sequence evolutionary information~\cite{Lin2023-jd, Jumper2021-xa}. The study of PLIs may eventually outgrow NLP methods, but for the foreseeable future, advances in NLP will continue to have a significant impact. Collaborations between experts in biological and computational domains will be critical for catalyzing further innovations in what is an interdisciplinary goal. 

\section{7. Conclusion}
Natural language processing (NLP), a sub\=/discipline of machine learning (ML), offers new tools for both experimental and computational researchers to accelerate exploratory studies in structural biology. The prediction of protein\=/ligand interactions (PLIs) can be re\=/imagined through NLP by treating protein and ligand representations like text. Protein sequences resemble readable text with inherent meaning to be inferred, while the SMILES format for chemical compounds allows limited NLP application to small molecules. Current efforts seek to leverage multiple or augmented SMILES representations to address these limitations. 

Approaches to tackling PLI prediction tasks using sequence\=/only data, structural data, or a combination of both, have all yielded successful predictions, although the advantage of one input data type over others remains unclear.  Sequence\=/only data approaches are simple and amenable to NLP but requires a significant abstraction of chemical information; structural data is informationally rich but computationally expensive to handle, while combining both sequence and structural data types offers balance at the expense of complexity.

The transformer architecture, in general, and attention mechanisms, in particular, have yielded the most promising NLP-based PLI prediction results to date. Incorporating complementary data (e.g., multiple sequence alignments, ligand polarities, etc.) can improve predictive success but at a significant increase in computational cost. After data selection and preparation, all methods have followed a general ML Extract\=/Fuse\=/Predict model creation framework of: (i) extracting feature embeddings for protein and ligand, (ii) fusing protein and ligand embeddings, and (iii) making predictions based on the created ML model. 

The first step of dataset selection is crucial for any ML-based study of PLIs, and no single dataset can satisfy all needs, with many suffering from missing data or lack of negative data. Datasets must align with specific research goals, requiring thoughtful consideration as to what inputs, formats, and target variable(s) are selected for the ML model. Appropriate tokenization and embedding methods, which convert proteins and ligands into numerical representations, are vital for a successful model. Atoms or amino acids typically serve as tokens, and neural networks (NNs) have helped to identify hidden patterns more quickly. NLP\=/inspired NNs, such as Long Short\=/Term Memory NNs, along with attention mechanisms and transformer architectures, have shown particular promise for understanding PLIs. A modular approach combining multiple embeddings can capture diverse perspectives, improving prediction accuracy, especially for the prediction of binding affinities. After appropriate embeddings are obtained, graph\=/based methods and cross\=/attention mechanisms have been shown to be effective in combining data from diverse sources.

NLP has been central to ML studies of PLIs and has yielded promising results, although many challenges remain. Explaining ML model predictions is essential for their trustworthiness and acceptance. Current explanatory metrics, such as attention weights and Shapley values, offer some degree of interpretability but remain to be fully validated. A major challenge is the lack of well\=/annotated non\=/binding protein\=/ligand pairs, or "negative data". Unsupervised methods or manually curated selections of non\=/binding pairs are potential solutions. Popular PLI datasets may contain biases that cause models to "memorize" idiosyncratic patterns rather than "learn" the true mechanics of PLIs. Ensuring balanced training datasets (positive vs. negative data, number of proteins vs. ligands, etc.) would be essential to avoid such bias.

As protein and ligand sequence representations differ from human language, it may be difficult to capture their complexity with NLP methods alone, especially as much of the variation in protein function can often be explained by simple amino acid interactions rather than complex higher\=/order interactions~\cite{Park2024-oj}. While NLP has contributed significantly to advancing our study of PLIs, future improvements may come from both modifying machine learning architectures and incorporating nuanced biological domain knowledge. For instance, the researchers behind AlphaFold\=/Multimer's protein-protein interaction prediction algorithm~\cite{Evans2021-ah} created an interface-aware protocol that crops protein structures to reduce computational burden and the representation of non-interfacial amino acids while maintaining an important balance of interacting and non-interacting regions. Some researchers have also integrated mass spectrometry data to improve model predictions of protein complexes~\cite{Stahl2024-go}. More recently in AlphaFold3~\cite{Abramson2024-qh}, a diffusion layer has been added to Alphafold’s previous workflow to enable the study of PLIs. Time will tell to what degree AlphaFold3 will advance predictions of PLIs but progress in PLI research will undoubtedly require interdisciplinary collaborations between computer scientists, chemists, and biologists.

Although it is best practice to evaluate model performance against ground\=/truth experimental results or results from physics\=/based computer simulations, few studies to date have benchmarked their model predictions in this way. Formal competition may prove to be a promising avenue for future advances in PLI prediction. Other grand challenges, such as protein folding and protein assembly, have had significant progress facilitated through competitions like Critical Assessment of Structural Prediction (CASP)~\cite{Kryshtafovych2019-gl} and Critical Assessment of Prediction of Interactions (CAPRI)~\cite{Janin2003-as,Lensink2020-fx}. These well\=/adjudicated competitions use unpublished test sets for objective model comparisons. Milestone algorithms like AlphaFold~\cite{Senior2020-fo} and RosettaFold~\cite{Baek2021-xl} were formed, improved, and refined through the crucible of such contests. Creating a dedicated competition devoted to protein-ligand interactions could similarly inspire innovation and catalyze seminal algorithmic advances for PLI prediction.

\begin{table}[]
\scriptsize
\centering
\caption{Glossary of Terms That Appear in the Tables}
\begin{tabularx}{\textwidth}{p{3.7cm} p{12cm}}
\hline
\textbf{Term} & \textbf{Definition} \\ \hline
\rowcolor{lightgray}
\textbf{AutoEncoder} & A neural network tasked with compressing and reconstructing input data, often used for feature learning.~\cite{Kramer1991-ou} \\
\textbf{Dilated Convolutional Block} & Convolutional Neural Network operations with defined gaps between kernels, which can capture larger receptive fields with fewer parameters.\\
\rowcolor{lightgray}
\textbf{ECFC4 Fingerprint} & A molecular fingerprint that encodes information about the presence of specific substructures within a diameter of 4 bonds from each atom.~\cite{Rogers2010-me} \\
\textbf{FCN} & Fully-Connected Network, a feedforward Neural Network where each neuron in one layer connects to every layer in the next. FCNs can also be referred to as Multi-Layer Perceptrons.\\
\rowcolor{lightgray}
\textbf{GAN Discriminator} & An NN part of Generative Adversarial Networks (GAN) that learns important features to distinguish between real and artificial data.\\
\textbf{GAT} & Graph Attention Network, a type of Graph Neural Network that uses attention mechanisms to deciding the value of neighboring nodes to a given node when updating a node's information.~\cite{Velickovic2017-pl}\\
\rowcolor{lightgray}
\textbf{GCN} & Graph Convolutional Network, a type of Graph Neural Network that aggregates neighboring node features through a first-order approximation on a local filter of the graph.~\cite{Kipf2016-bo} \\
\textbf{GIN} & Graph Isomorphism Network, a type of Graph Neural Network that uses a series of functions to ensure embeddings are the same no matter what order nodes are presented in.~\cite{Xu2017-em} \\
\rowcolor{lightgray}
\textbf{Gradient-Boosting Trees} & A machine learning technique where many decision trees are trained in order, such that the next tree learns from the misclassified samples of the previous tree. All trees are then used to "vote" on results of each input.\\
\textbf{GRU} & Gated Recurrent Unit, a simplified version of Long Short-Term Memory that similarly uses a gating mechanism to retain and forget information, but is less complex than Long Short-Term Memory.~\cite{Cho2014-xe} \\
\rowcolor{lightgray}
\textbf{Heterogeneous Network} & A graph where nodes and edges represent different types of information, often used to convey complex relationships in biological systems (e.g. drug, target, side-effect, etc.).\\
\textbf{Message-Passing NN} & Type of Graph Neural Network that computes individual messages to be passed between nodes so that representations for each node contain information from its neighbors~\cite{Gilmer2017-ga}.\\
\rowcolor{lightgray}
\textbf{ProtVec} & A method for representing protein sequences as dense vectors using skip-gram neural networks.~\cite{Asgari2015-eb}\\
\textbf{Random Forest} & A machine learning method where many decision trees are constructed, and the result of the ensemble is the mode of the individual tree predictions.\\
\rowcolor{lightgray}
\textbf{ResNet} & Short for Residual Network. A neural network architecture that speeds up training by learning functions to substitute for layer operations, allowing for the "skipping" of layers and faster training.~\cite{He2015-ou}\\
\textbf{Seq2Seq} & A machine learning method used for language translation in NLP, featuring an encoder-decoder structure.~\cite{Xu2017-em}\\
\rowcolor{lightgray}
\textbf{SMILESVec} & Previous work from authors. 8-character ligand SMILES fragments are assigned a vector through a single-layer neural network, and an input SMILES string's vector is equal to the mean of fragment vectors present in that input SMILES.~\cite{Ozturk2018-zg}\\
\textbf{Squeeze-And-Excite Mechanism} & Mechanism for Convolutional Neural Networks that uses global information to adapt the model to emphasize more important features.~\cite{Hu2018-bl} \\
\bottomrule
\end{tabularx}
\label{Glossary}
\end{table}

\begin{acknowledgement}

This work was supported in part by NIGMS/NIH Institutional Development Award (IDeA) \#P20GM130460 to J.L, NSF award \#1846376 to E.F.Y.H, and University of Mississippi Data Science/AI Research Seed Grant award \#SB3002 IDS RSG-03 to J.M., J.L., T.L, and E.F.Y.H. 

\end{acknowledgement}

\bibliography{paperpile}

\newpage
\begin{figure}[h!]
   \centering      
   \includegraphics[]{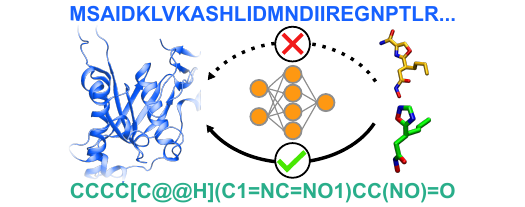}     
 \caption*{Table of Contents Graphic}
\end{figure}

\end{document}